\documentstyle[12pt,epsfig]{article}

\graphicspath{{ridge-img/}}

\begin{document}
\begin{center}
{\large \bfseries Onset of "ridge phenomenon" in $AA$ and $pp$ collisions and percolation string model}
\vskip 5mm

O.Kochebina and G.Feofilov

{\small
{\it V.Fock Institute for Physics of Saint-Petersburg State University}
\\
$\dag$ {\it
E-mail: feofilov@hiex.niif.spb.su, kochebina@gmail.com
}}
\end{center}

\vskip 5mm

\begin{center}

\begin{minipage}{150mm}
\centerline{\bf Abstract}
Detailed study of centrality dependence of low-$p_t$ manifestation of so-called "near-side ridge" phenomenon is reported recently by STAR for all charged
hadrons with $p_{t} > 0.15 GeV/c$ from $AuAu$ collisions at 62 and 200~GeV at RHIC~\cite{Daugherity}. It is indicating the existence of the
energy-dependent centrality point where some sudden changes in the correlation pattern are observed. In the present work we use the hypothesis of string percolation phase transition for the description of the onset of this ridge. One may assume that the formation of rather large "macroscopic" clusters composed of several overlapped strings extended in rapidity and localized in azimuth could be one of the possible processes leading to the observed  phenomenon.

This onset is characterized by some definite ("critical") number of participating nucleons ($N_{part}^{crit}(\sqrt{s})$). We use also another
physical quantity, transverse particle density ($\tilde{\rho}$), to characterize this threshold behavior, this variable brings the transition points for two energies to coincidence~\cite{Daugherity}. So parameters of the percolation model are defined at these critical points.
Also we use results of our previous calculations~\cite{j-psi} for energy of collision $\sqrt{s}$~=~17.3~GeV based on observed threshold of anomalous suppression of $J/\psi$.
 Obtained parameters are extrapolated to $AA$ and $pp$ collisions for energies over the range 17.3~GeV -- 7000~GeV.
\end{minipage}
\end{center}

\section{Introduction and motivation.}

Dihadron correlations that were studied in $AuAu$ collisions at the collision energies $\sqrt{s}$ = 62 and 200~GeV, demonstrate a number of interesting features as a function of collision centrality, transverse momentum and particle composition~\cite{Putschke,STAR 2006,STAR 2007}. Results obtained at RHIC indicate that in addition to the jet-like component, the long-range pseudorapidity azimuthal correlations are observed in $AuAu$
collisions forming a so-called ridge structure. Several theoretical models were proposed to explain qualitatively the ridge origin using
the various concepts that are in one or another way relevant to the interactions of high-$p_{t}$ partons or jets with medium, jets in small-$p_{t}$~\cite{Capella},
parton-jets collisions etc, see references in~\cite{Putschke}.

\begin{figure}[h]
\begin{center}
\includegraphics[width = 350pt, height = 180pt]{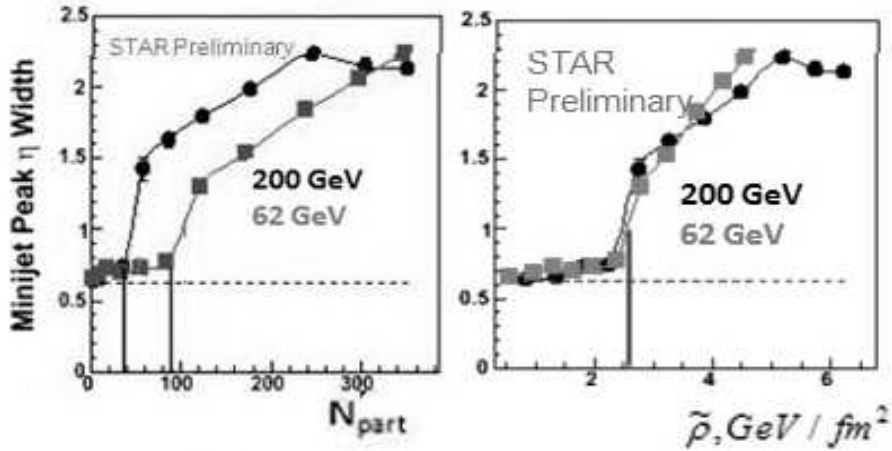}
\caption{Minijet peak width in $\eta$ vs. the number of nucleon-participants $N_{part}$ (left figure) and vs. transverse particle density  $\tilde{\rho}$ (right figure) measured for $AuAu$ collisions at two collision energies of 62 and 200 GeV~\cite{Daugherity}. Vertical lines are marking "critical" values of $N_{part}$ and $\tilde{\rho}$.
}
\label{ris:ro}
\end{center}
\end{figure}

However, recently, the experimental ridge landscape was broadened by the detailed study of energy and centrality dependence of the minijet angular correlations measured for all charged hadrons with
 rather low-$p_{t}$ ($p_{t} > 0.15 GeV/c$) in $AuAu$ collisions at 62 and 200~GeV in STAR at RHIC~\cite{Daugherity}.
The data presented in~\cite{Daugherity} display jet-like correlations which change dramatically with centrality~(Fig.\ref{ris:ro}), -- a dependence which was not anticipated by theoretical calculations of parton energy loss and medium modified fragmentation based on perturbative quantum chromodynamics (pQCD) jet-quenching models~\cite{Wang-Gyulassy, Armesto-Flow} or  parton recombination models~\cite{Hwa}.

These new data showed a sharp transition at approximately 55\% centrality for 200~GeV and at about 40\% for 62~GeV collision energy, where the near-side amplitude and pseudorapidity widths of the ridge increased dramatically while the angular widths continued to decrease slightly~\cite{Daugherity}(see Fig.\ref{ris:ro}). The onset of this transition is marked by some definite ("critical") number of participating nucleons ($N_{part}^{crit}(\sqrt{s})$).
One may see from the data~\cite{Daugherity} that at the collision energy 62~GeV this phenomenon starts at $N_{part}^{crit}\sim90$, and at the energy 200~GeV the relevant threshold is marked by  $N_{part}^{crit}\sim40$ (see Fig.\ref{ris:ro}).

It was also found in~\cite{Daugherity} that the transverse particle density
\begin{equation}
\label{rho}
\tilde{\rho}=\frac{3}{2}~\frac{\frac{dN_{ch}}{dy}}{<S>}
\end{equation}
brings the transition points for these two energies to coincidence at the value $2.6\pm0.2$~GeV/fm$^{2}$ (Fig.\ref{ris:ro}, right).
Here $\frac{dN_{ch}}{dy}$ -- is the charge particle multiplicity per rapidity unit, $<S>$ -- the
 collision overlap area, the factor $\frac{3}{2}$ appears because both charge and neutral particles are taken in account. It is important to note that both $\frac{dN_{ch}}{dy}$ and $<S>$ in Eq.(\ref{rho}) depend on the number of nucleons-participants $N_{part}$.

This low-$p_{t}$ manifestation of the "ridge" structure, expanding in pseudorapidity \cite{Daugherity}, stimulated our present attempt to consider the onset of this phenomenon on the base of the string percolation model. The basic parameters of this model are defined below by using the experimental data~\cite{Daugherity} on the centrality dependence
observed for the minijet angular long-range correlations.

\section{Critical string density and onset of low-$p_{t}$ ridge.}
Colour strings~\cite{Kaidalov,Armesto2000} may be viewed as tubes of the colour field created by the colliding partons. The creation of particles goes via spontaneous emission of quark-antiquark pairs in this colour field. These strings are the phenomenological objects extended in rapidity, their cross section in the transverse plane is considered to be small discs of $\pi r_{0}^2$ area, where $r_{0}$ is the string radius. With growing energy and/or atomic number of colliding particles the number of 
strings grows, and they start to overlap forming clusters. The last statement is true only in case of the interacting strings~\cite{Abramovsky}. Therefore, at some definite critical string density the percolation phase transition could be reached and new particle emitting sources could appear~\cite{Armesto1996,Braun,Nardi,Braun2000}. As it was pointed in~\cite{Armesto1996,Nardi,Braun2000}, the percolation of strings can be considered as a smooth way to quark gluon plasma after the cluster thermalization. It was also noted in~\cite{Kharzeev} that the string percolation phenomenon is closely related to the parton saturation in high density QCD.

We assume in the present study that it is the formation of these rather large "macroscopic" clusters composed of several overlapped strings, extended in rapidity and localized in azimuth, that is leading to the observed near-side ridge phenomenon.
\subsection{String density in $AA$ collisions}
To characterize the
string density in the transverse overlap area of colliding nuclei a dimensionless percolation parameter $\tilde{\eta}$ was introduced~\cite{Armesto1996,Nardi,Braun2000}:

\begin{equation}
\label{perc}
\tilde{\eta}=\frac{\pi r_{0}^2 N_{str}}{<S>}
\end{equation}

Here $<S>$ is the mean transverse area of nuclear interaction (overlap area) at a given number of nucleons-participant,
$N_{str}$ is a number of strings that is also relevant to the given $N_{part}$. 
The critical value of the parameter $\tilde{\eta}$ marking the percolation transition ($\tilde{\eta}^{crit}$) could be calculated from the geometrical considerations.
Generally "critical" values of $\tilde{\eta}^{crit}$ are used to be $\approx~1.12-1.175$~\cite{BraunPRL}, and string radius
is  usually taken as $r_{0}=0.2-0.3$~fm~\cite{Armesto2000, Dias de Deus, Pajares}. In our calculations below we use $\tilde{\eta}^{crit}~\approx~1.15\pm0.03$ and $r_{0}~=~0.25$~fm.

The number of particle emitting sources - strings - $N_{str}$, produced in the nucleus-nucleus collision, generally depends on the centrality of collision, on the type of colliding systems and on the collision energy $\sqrt{s}$.




 The overlap area $<S>$ of Eq.(\ref{rho}) and Eq.(\ref{perc}) is not precisely defined.
 We propose to exclude $<S>$ from the estimations by considering the ratio
$\tilde{\rho}^{crit}/\tilde{\eta}^{crit}$ (see~Eq.(\ref{rho})~and~Eq.(\ref{perc})) at the "critical" point, that  marks the onset of the low-$p_{t}$ ridge manifestation mentioned above.
It is evident that the dependence on the variable $<S>$ could be excluded in this case.
Thus one can obtain at the critical point:

\begin{equation}
\label{ratio}
\frac{\tilde{\rho}^{crit}(N_{part})}{\tilde{\eta}^{crit}(N_{part})}=\frac{3}{2} \frac{1}{\pi r_{0}^2} \frac{dN_{ch}}{dy} \frac{1}{N_{str}} = 2.3 \pm 0.2 GeV/fm^{2}
\end{equation}

where error is coming mainly from the systematic uncertainties of $\tilde{\rho}^{crit}$ and $\tilde{\eta}^{crit}$.
The total number of the strings $N_{str}$ could be easily found from the Eq.(\ref{ratio}) for the given $AA$ collision energy and at the given "critical" centrality ($N_{part}^{crit}$). That means that parameters of percolation model defining the total number of strings $N_{str}$ can be obtained. So, it is also possible to find further the energy and centrality dependence of string density by using the available data.

\subsection{Centrality and energy dependence of $\tilde{\rho}$ and $\tilde{\eta}$}

In order to study the energy and centrality dependence of string density parameter $\tilde{\eta}$ we use in the present work
the concept of valence and sea strings, so that:
\begin{equation}
\label{valent-sea}
N_{str}=N_{V}+N_{S}
\end{equation}
Here $N_{V}$ is number of the strings formed by the valence quarks and $N_{S}$ is number of the strings from the sea quarks. $N_{V}$ is defined by the number of nucleons-participants $N_{part}$, and $N_{S}$ here depends on $N_{coll}$ -- the number of nucleon-nucleon collisions relevant to the given $N_{part}$.

We propose to use the following parametrization:
\begin{equation}
\label{a-prop}
N_{str}=N_{part}+aN_{coll}
\end{equation}
That means that the number of the sea strings $N_{S}$ formed in $AA$ interaction is proportional to the number of nucleon-nucleon collisions $N_{coll}$.
The coefficient $a$ could be defined from $N_{str}$ estimated by using Eq.(\ref{ratio}) at the "critical" point.

The energy dependence enters into the Eq.(\ref{a-prop}) via $a$ and $N_{coll}$.
Values of $N_{coll}$ relevant to the given $N_{part}$ at Eq.(\ref{a-prop}) are obtained from the Monte Carlo simulations of the Modified Glauber model(MGM)~\cite{Ivanov}. The last one takes in account nucleon momentum loss during collision and describes the centrality and energy dependence of charged particle multiplicity measured in relativistic $AA$ collisions~\cite{Ivanov}.

So the parameter $a$ is obtained from Eq.(\ref{ratio}) and Eq.(\ref{a-prop}) using the values of $\frac{dN_{ch}}{dy}$ taken from the known experimental data.

One can find in the Table~\ref{tab:results-AA} the results of calculation of parameter $a$ for energies 62 and 200~GeV. Also we use results of our previous calculations for 17.3~GeV~\cite{j-psi} based on observed threshold of anomalous suppression of $J/\psi$.
\begin{table}[h]
\begin{center}
\begin{tabular}{|c|c|c|c|c|c|c|c|c|c|}
 \hline
$\sqrt{s},~GeV$ & $N_{part}$ & $dN_{ch}/d\eta$ & $N_{str}$ & $N_{s}$ & $N_{coll}$ & $a$ \\
 \hline
200($AuAu$) &  40 &  $2.97\pm0.30$~\cite{PHOBOS130GeV-200GeV} &  $194\pm25$ & $155\pm23$&  $59\pm4$ & $2.6\pm0.4$ \\
62 ($AuAu$)&  90 &  $2.30\pm0.23$ ~\cite{PHOBOS62} &  $352\pm28$ & $262\pm23$ &  $167\pm4$ & $1.6\pm0.2$\\
17.3 ($PbPb$)&  110 &  $1.62\pm0.21$~\cite{NA57-17.3} & $302\pm45$ & $192\pm30$ &  $158\pm5$& $1.2\pm0.2$ \\
 \hline
\end{tabular}
\caption{The total number of strings, the number of sea strings and parameter $a$ obtained in the "critical" points for $AuAu$ collisions at 62~GeV and 200~GeV collision energies and for $PbPb$ collisions at 17.3~GeV. The calculations are done for $r_{0}=0.25$~fm.}
\label{tab:results-AA}
\end{center}
\end{table}
\begin{figure}[h]
\begin{center}
\includegraphics[width = 250pt, height = 140pt]{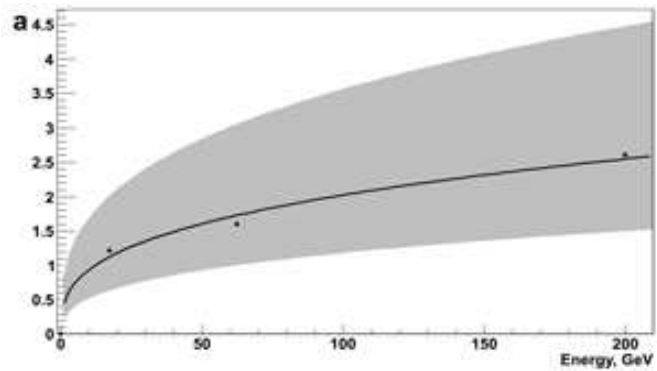}
\caption{Energy dependence of parameter $a$ (Eq.(\ref{a-prop})).}
\label{ris:a-energy}
\end{center}
\end{figure}

It is possible to make some extrapolation of parameter $a$ to lower and higher energies (Fig.\ref{ris:a-energy}). In case of lower energies it can be done straightforwardly to the point (0;0).
Extrapolation could be also prolonged to the higher energies with some uncertainties -- by fitting the data with some suitable analytic curve.
\begin{equation}
\label{alpha}
a=A+B(\sqrt{s})^{(1/3)}
\end{equation}
The corresponding plot is presented on Fig.\ref{ris:a-energy} with parameters of extrapolation $A~=~0.039$ and $B~=~0.13$. It is evident that the error bars are currently too large, so the extrapolation is rather rough. Nevertheless it is possible to estimate further the total number of strings for other energies and to get centrality dependence of string density $\tilde{\eta}$ for different colliding systems.

For getting the centrality dependence of $\tilde{\eta}$ we are forced again to introduce into the calculations the  interaction area $<S>$. The last one may be estimated by applying the relation $<S>\sim~N_{part}^{2/3}$~\cite{S_Npart}. The coefficient of proportionality here is derived by using the information obtained at the "critical" point of the transverse particle density. Modified Glauber model~\cite{Ivanov} is used again here for getting $N_{part}$ and $N_{coll}$.

\begin{figure}[h]
\begin{center}
\includegraphics[width = 320pt, height = 247pt]{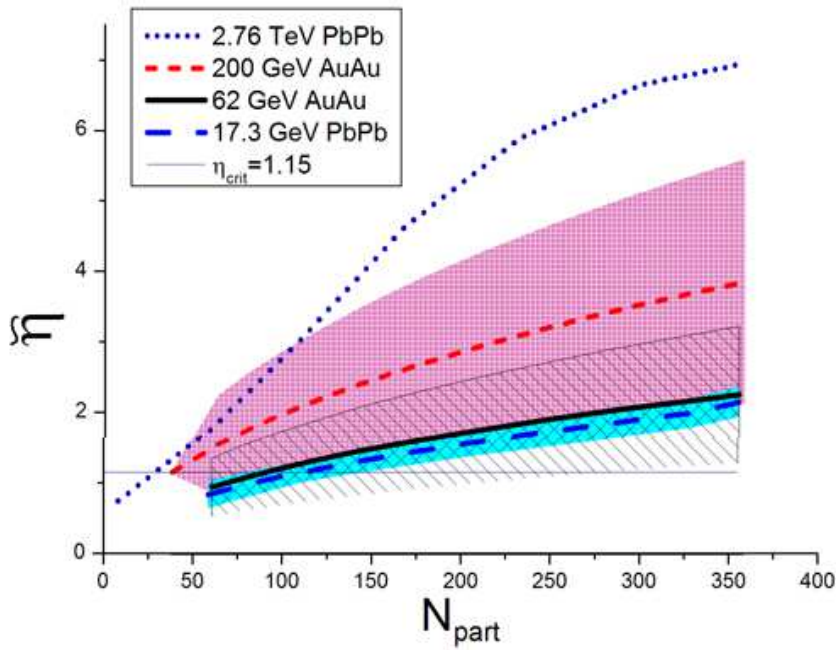}
\caption{Centrality dependence: parameter $\tilde{\eta}$ vs. number of participants $N_{part}$ in $AA$ collisions at various energies. Shaded areas are representing the uncertainties of calculations.
}
\label{ris:centrality}
\end{center}
\end{figure}

The obtained centrality dependence of percolation parameter for $AA$ collisions is shown on Fig.\ref{ris:centrality} for energies $\sqrt{s}$~=~17.3~GeV, 62~GeV, 200~GeV and 2.76~TeV per nucleon-nucleon pair. Points for the LHC energies are derived by using theoretical predictions for $\frac{dN_{ch}}{dy}$~\cite{Armesto2005}.
Uncertainties at Fig.\ref{ris:centrality} come from the calculation of $N_{part}$, $N_{coll}$ and from the uncertainties of definition of the "critical" values of $\tilde{\rho}$ and $\tilde{\eta}$ and of $r_{0}$. The main contribution is from the systematic uncertainties of $\tilde{\rho}_{crit}$, $\tilde{\eta}_{crit}$ and $r_{0}$. Extrapolation to the $PbPb$ collisions at center-of-mass energy per particle pair 2.76~TeV, that holds also error of fitting Fig.\ref{ris:a-energy}, contains very large uncertainties of the order of 40\% for central collisions. They are not shown on Fig.\ref{ris:centrality} to keep readability.

One may see on the Fig.\ref{ris:centrality} that rather large values of average string density $\tilde{\eta}$ might be expected at LHC energies in the central $PbPb$ collisions up to 4-6 and in the very peripheral collisions, where $N_{part}\approx50$ -- up to 1.5.
So one could search the ridge phenomenon in all classes of centrality.

These calculations provide the possibility to study the system size and energy dependence of the percolation parameter and to search further the critical density effects.

\section{Transverse particle density and percolation of strings  in $AA$ and $pp$ collisions.}

Thus we have two densities: from one side the transverse particle density based on experimental data~\cite{PHOBOS130GeV-200GeV, PHOBOS62, NA57-17.3}, and on the other side -- the estimated density of strings in percolation string model in $AA$ collisions at different centralities and collision energies. From Eq.(\ref{ratio}), we can get a formal relation between $\tilde{\rho}$ and $\tilde{\eta}$:

\begin{equation}
\label{ro-eta}
\tilde{\rho}^{crit}(N_{part})=\frac{3}{2} \frac{1}{\pi r_{0}^2} \frac{dN_{ch}}{dy} \frac{1}{N_{str}} \tilde{\eta}^{crit}(N_{part})
\end{equation}
We have to note here that Eq.(\ref{ro-eta})
could be non linear due to the behavior of $\frac{dN_{ch}}{dy}$ and $N_{str}$ with centrality and collision energy.
In our work the correspondence of the transverse particle density and density of the strings is being set at the critical point. Then the extrapolation to other systems and energies is done.

\begin{figure}[h]
\begin{center}
\includegraphics[width = 320pt, height = 247pt]{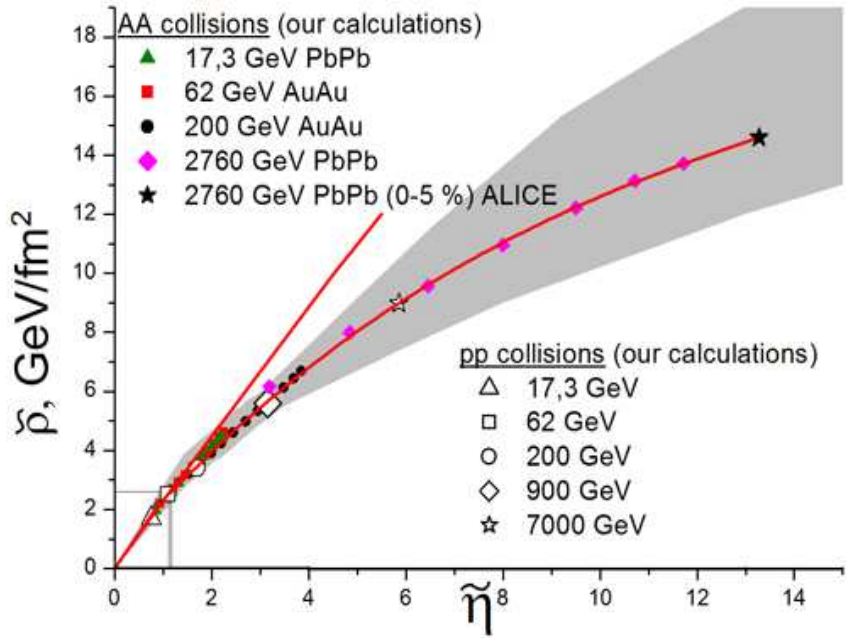}
\caption{
Transverse particle density $\tilde{\rho}$ vs. string density $\tilde{\eta}$ in $AA$ collisions for energy $\sqrt{s}$~=~17.3~GeV, $\sqrt{s}$~=~62, $\sqrt{s}$~=~200~GeV and $\sqrt{s}$~=~2.76~TeV. Points are our calculations. Lines are fits to the points. Points for $pp$ collisions are obtained from the experimental data in the range of energies from $\sqrt{s}$~=~17.3~GeV to $\sqrt{s}$~=~7~TeV (see text).}
 \label{ris:eta-ro}
\end{center}
\end{figure}

The filled points on the Fig.~\ref{ris:eta-ro} are the results of our calculations of the transverse particle density $\tilde{\rho}$ vs. string density $\tilde{\eta}$ in $AA$ collisions
 for different energy values of $\sqrt{s}$~=~17.3 GeV, $\sqrt{s}$~=~62~GeV, $\sqrt{s}$~=~200~GeV and $\sqrt{s}$~=~2.76~TeV.
Lines are fits to the points: linear fit is used before the critical point ( $y~=~0.02+2.28\cdot~x$) and the exponential one after (by the curve $y~=~19.5-19.1\cdot~e^{-x/9.9}$). This deviation from the linear behavior after the critical point can be explained by the fact that after the percolation threshold a new types of particle emitting sources appear. Thus the transverse particle density may be  decreased,  which is in line with the prediction of string percolation model where the yield of multiplicity should be reduced in case of string fusion~\cite{Braun2000}.

The empty points on the Fig.~\ref{ris:eta-ro} correspond to $pp$ collisions. In case of $pp$ collisions we assume that dependence of $\tilde{\rho}$ -- $\tilde{\eta}$ is the same as in the case of $AA$ collisions. The concept of valent and sea strings is valid, but it is evident that the parametrization Eq.(\ref{a-prop}) can not be used here.
The transverse particle density in $pp$ collisions is calculated on the base of the experimental data~\cite{Exp7000GeV, Exp900GeV, Exp200GeVp-antip, Exp62GeV, Exp17GeV} of the charge particle multiplicity per rapidity unit and assuming that the interaction area equals to $1~fm^{2}$. Thus the relevant values of $\tilde{\eta}$ could be obtained from the Fig.\ref{ris:eta-ro}, and one can also get the total number of strings in the interaction region. Results are summarized in the Table~\ref{tab:result}.

\begin{table}[h]
\begin{center}
\begin{tabular}{|c|c|c|c|c|c|c|c|c|}
 \hline
 $\sqrt{s},~GeV$ &  $dN_{ch}/d\eta$ &  $\tilde{\rho},~GeV/fm^{2}$ &  $\tilde{\eta}$ &  $N_{str}$ our& $N_{str}$ estimate & Dispersion in \\
 &  &  &  & estim. & using data~\cite{derkach}& $N_{str}$ from ~\cite{derkach} \\
 \hline
$7000$ &  $6.02\pm0.50$~\cite{Exp7000GeV} &  $9.0\pm0.5$ & $5\pm1$ &  $30\pm12$ & no data & no data \\
900 & $3.78\pm0.19$~\cite{Exp900GeV} & $5.6\pm0.4$ &  $3.2\pm0.8$ &  $16\pm6$ & 6.6 & 8.6 \\
200($p\bar{p}$) &  $2.30\pm0.20$~\cite{Exp200GeVp-antip} &  $3.4\pm0.2$ &  $1.7\pm0.4$ &  $9\pm4$ & 5 & 5.4 \\
62 &  $1.64\pm0.17$~\cite{Exp62GeV} &  $2.5\pm0.1$ &  $1.1\pm0.6$ &  $6\pm4$ & 4 & 3.6\\
17.3 &  $1.14\pm0.12$~\cite{Exp17GeV} &  $1.70\pm0.08$ &  $0.7\pm0.4$ &  $4\pm3$ & 3 & 1\\
 \hline
\end{tabular}
\caption{The transverse particle density, the string density and number of strings estimated for $pp$ collisions for energies from 17.3~GeV to 7000~GeV.}
\label{tab:result}
\end{center}
\end{table}

In the  Table~\ref{tab:result} we compare our estimates with those done independently in the framework of the multipomeron exchange model~\cite{derkach}. The number of strings is considered as the number of pomeroms multiplied by factor 2 ($N_{str}~=~N_{pomerons}~\times~2$) following the assumption that one cut pomeron gives two strings. We have to note that in our calculations there are rather large uncertainties, those come from estimation of $N_{part}$, $N_{coll}$, the systematic uncertainties of definition of the "critical" values of $\tilde{\rho}$ and $\tilde{\eta}$ and of $r_{0}$ and from extrapolation of parameter $a$ on Fig.\ref{ris:a-energy}. However, quite satisfactory agreement with the results that are based on this different calculation approach could be noted. So, in case of the $pp$ collisions at the energies more than $\approx 65$~GeV one could expect the density of  strings to be  already sufficiently high enough to reach the percolation phase transition threshold  and to produce the long-range correlations.

We also show on the Fig.~\ref{ris:eta-ro} by the filled star the estimation for the central $PbPb$ collisions at ALICE at $\sqrt{s}$~=~2.76~TeV based on the first experimental data, where $\frac{dN_{ch}}{d\eta}~=~8.3\pm~0.4~(sys.)$~\cite{ALICEPbPB}.

\section{Conclusions.}
\begin{enumerate}

  \item The hypothesis of percolation transition looks reasonable in the description of the onset of the low-$p_{t}$ manifestation near side  ridge phenomena in $AuAu$ collisions at RHIC energies. One may assume that at sufficiently high string densities the formation of rather large "macroscopic" clusters composed of several overlapped strings extended in rapidity and localized in azimuth could be one of the possible processes leading to the observed  ridge phenomenon.

  \item It is shown that in $PbPb$ collisions at the LHC energies the onset of the  near-side low-$p_{t}$ manifestation of "ridge" should be observed in  all classes of centrality

      \item The new method to estimate the total number of sea strings formed in the AA collisions is proposed.

  \item Our estimates show that sufficiently high string density and the appearance of the long-range azimuthal correlations of charged particles (including ridge) could be also expected in $pp$ collisions at RHIC energies. One may expect very high overlap of strings (about 5) in $pp$ collisions at the LHC.

\end{enumerate}

\section{Acknowledgements.} Authors would like to thank V.~Vechernin and G.~Paic for fruitful discussions and for interest to this work. These studies were partially supported for Olga Kochebina by the grant of the non-profit Dynasty Foundation (Russian Federation).

\end{document}